\documentclass[10pt,letter]{IEEEtran}

\usepackage{amssymb,calc,epsfig,graphicx,psfrag}
\begin{document}
\newtheorem{claim}{Claim}
\newtheorem{lemma}{Lemma}
\newtheorem{theorem}{Theorem}
\newtheorem{definition}{Definition}
\newtheorem{example}{Example}
\newtheorem{conjecture}{Conjecture}

\newcommand{\enp} {\hfill \rule{2.2mm}{2.6mm}}
\newcommand{\dk} {\preceq_k}
\newcommand{\sm} {\equiv}
\newcommand{\lf}{\rightrightarrows}
\newcommand{\lfe}{\rightarrow}
\newcommand{\ints}{{\cal Z}}
\newcommand{\Mx}{\mbox{Mx}}

\onecolumn

\title{Maximum Weight Matching via Max-Product Belief Propagation}


\author{Mohsen Bayati ~~ Devavrat Shah ~~ Mayank Sharma\thanks{M. Bayati
is with Department of EE at Stanford University; D. Shah is with Departments
of EECS and ESD at MIT and M. Sharma is with IBM T.J. Watson Research Center.
Email:  bayati@stanford.edu, devavrat@mit.edu, mxsharma@us.ibm.com.}}


\maketitle

\begin{abstract}

Max-product ``belief propagation"  is an iterative, local,
message-passing algorithm for finding the maximum a posteriori (MAP)
assignment of a discrete probability distribution specified by a
graphical model. Despite the spectacular success of the algorithm in
many application areas such as iterative decoding, computer vision
and combinatorial optimization which involve graphs with many
cycles, theoretical results about both correctness and convergence
of the algorithm are known in few cases \cite{WeissFreeman,
wainwright04, YFW00-22,Urbanke}.

In this paper we consider the problem of finding the Maximum
Weight Matching (MWM) in a weighted complete bipartite graph.  We
define a probability distribution on the bipartite graph whose MAP
assignment corresponds to the MWM.  We use the max-product
algorithm for finding the MAP of this distribution or
equivalently, the MWM on the bipartite graph.  Even though the
underlying bipartite graph has many short cycles, we find that
surprisingly, the max-product algorithm always converges to the
correct MAP assignment as long as the MAP assignment is unique. We
provide a bound on the number of iterations required by the
algorithm and evaluate the computational cost of the algorithm.
We find that for a graph of size $n$, the computational cost
of the algorithm scales as $O(n^3)$, which is the same as the
computational cost of the best known algorithm. Finally, we
establish the precise relation between the max-product algorithm
and the celebrated {\em auction} algorithm proposed by Bertsekas.
This suggests possible connections between dual algorithm and max-product
algorithm for discrete optimization problems.

\end{abstract}

\section{INTRODUCTION}

Graphical models (GM) are a powerful method for representing and
manipulating joint probability distributions. They have found major
applications in several different research communities such as
artificial intelligence \cite{Pearl1988}, statistics
\cite{Lauritzen}, error-correcting codes \cite{Gallager1963, Horn,
Urbanke} and neural networks. Two central problems in probabilistic
inference over graphical models are those of evaluating the {\it
marginal} and {\it maximum a posteriori} (MAP) probabilities,
respectively. In general, calculating the marginal or MAP
probabilities for an ensemble of random variables would require a
complete specification of the joint probability distribution.
Further, the complexity of a brute force calculation would be
exponential in the size of the ensemble. GMs assist in exploiting
the dependency structure between the random variables, allowing for
the design of efficient inference algorithms.

The belief propagation (BP) and max-product algorithms
\cite{Pearl1988} were proposed in order to compute, respectively,
the marginal and MAP probabilities efficiently. Comprehensive
surveys of various formulations of BP and its generalization, the
junction tree algorithm, can be found in \cite{McEliece2000,
YFW00-22, Wainwright2003}. BP-based message-passing algorithms have
been very successful in the context of, for example, iterative
decoding for turbo codes, computer vision and finding satisfying
assignments for random k-SAT. The simplicity, wide scope of
application and experimental success of belief propagation has
attracted a lot of attention recently \cite{McEliece2000, Horn,
MPZ2002, Urbanke, YFW00-26}.

BP (or max-product) is known to converge to the correct marginal
(or MAP) probabilities on tree-like graphs \cite{Pearl1988} or
graphs with a single loop \cite{McEliece1998, Weiss2000}. For
graphical models with arbitrary underlying graphs, little is known
about the correctness of BP. Partial progress consists of
\cite{WeissFreeman} where the correctness of BP for Gaussian GMs
is proved, \cite{Koetter2000} where an attenuated modification of
BP is shown to work, and \cite{Urbanke} where the iterative turbo
decoding algorithm based on BP is shown to work in the asymptotic
regime with probabilistic guarantees. To the best of our
knowledge, little theoretical progress has been made in resolving the
question: Why does BP work on arbitrary graphs?

Motivated by the objective of providing justification for the
success of BP on arbitrary graphs, we focus on the application of
BP to the well-known combinatorial optimization problem of finding
the Maximum Weight Matching (MWM) in a bipartite graph, also known
as the ``Assignment Problem". It is standard to represent
combinatorial optimization problems, like finding the MWM, as
calculating the MAP probability on a suitably defined GM which
encodes the data and constraints of the optimization problem. Thus,
the max-product algorithm can be viewed at least as a heuristic
for solving the problem. In this paper, we study the performance of
the max-product algorithm as a method for finding the MWM on a
weighted complete bipartite graph.

Additionally, using the max-product algorithm for problems like
finding the MWM has the potential of being an exciting application of
BP in its own right. The assignment problem is extremely well-studied
algorithmically. Attempts to find better MWM algorithms contributed
to the development of the rich theory of network flow algorithms
\cite{EdmondsKarp1972, Lawler}. The assignment problem has been
studied in various contexts such as job-assignment in manufacturing
systems \cite{EdmondsKarp1972}, switch scheduling algorithms \cite{nick}
and auction algorithms \cite{Bertsekas}. We believe that the
max-product algorithm can be effectively used in high-speed switch
scheduling where the distributed nature of the algorithm and its
simplicity can be very attractive.

The main result of this paper is to show that the max-product
algorithm for MWM always finds the correct solution, as long as
the solution is unique. Our proof is purely combinatorial and
depends on the graph structure. We think that this result may lead
to further insights in understanding how BP algorithms work when
applied to other optimization problems. The rest of the paper is
organized as follows: In Section \ref{s:one}, we provide the
setup, define the Maximum Weight Matching problem (or assignment
problem) and describe the max-product algorithm for finding the
MWM. Section \ref{s:two} states and proves the main result of this
paper. Section \ref{s:minsum1} presents a simplification of the
max-product algorithm and evaluates its computational cost.
Section \ref{s:auction} discusses relation between the max-product
algorithm and the celebrate auction algorithm proposed by Bertsekas.
The auction algorithm essentially solves the dual of LP relaxation
for matching problem. Our result suggests possibility of deeper
connection between max-product and dual algorithm for optimization
problems. Finally, we discuss some implications of our results
in Section \ref{s:conc}.

\vspace{.05in}
\section{SETUP AND PROBLEM STATEMENT} \label{s:one}
\vspace{.05in}

In this section, we first define the problem of finding the MWM in a
weighted complete bipartite graph and then describe the max-product
BP algorithm for solving it.

\vspace{.05in}
\subsection{MAXIMUM WEIGHT MATCHING}
\vspace{.05in}

Consider an undirected weighted complete bipartite graph
$K_{n,n}=(V_1,V_2,E)$, where $V_1 = \{\alpha_1,\ldots,\alpha_n\}$,
$V_2=\{\beta_1,\ldots,\beta_n\}$ and $(\alpha_i,\beta_j) \in E$ for
$1 \leq i,j \leq n$. Let each edge $(\alpha_i,\beta_j)$ have weight
$w_{ij}\in \mathbb{R}$.

If $\pi=\{\pi(1),\ldots,\pi(n)\}$ is a permutation of
$\{1,\ldots,n\}$ then the collection of $n$ edges
$\{(\alpha_1,\beta_{\pi(1)}),\ldots,(\alpha_n,\beta_{\pi(n)})\} $ is
called a {\em matching} of $K_{n,n}$. We denote both the permutation
and the corresponding matching by $\pi$. The weight of matching $\pi$,
denoted by $W_{\pi}$, is defined as
\[
W_{\pi}=\sum_{1\leq i \leq n}w_{i\pi(i)}.
\]
Then, the Maximum Weight Matching (MWM), $\pi^*$, is the matching
such that
$$\pi^*=\textrm{argmax}_{\pi}\ W_{\pi}.$$

\vspace{.05in}

\noindent {\bf Note 1.} In this paper, we always assume that the
weights are such that the MWM is unique. In particular, if the
weights of the edges are independent, continuous random variables,
then with probability $1$, the MWM is unique.

\vspace{.05in}

Next, we model the problem of finding MWM as finding a MAP
assignment in a graphical model where the joint probability
distribution can be completely specified in terms of the product
of functions that depend on at most two variables (nodes). For
details about GMs, we urge the reader to see \cite{Lauritzen}.
Now, consider the following GM defined on $K_{n,n}$: Let
$X_1,\ldots,X_n,Y_1,\ldots,Y_n$ be random variables corresponding to
the vertices of $K_{n,n}$ and taking values from $\{1,2,\ldots,n\}$.
Let their joint probability distribution,
$p\left(\overline{X}=(x_1,\ldots,x_n);\overline{Y}=(y_1,\ldots,y_n)\right)$,
be of the form:
\begin{equation}
p\left(\overline{X},\overline{Y}\right)= \frac{1}{Z}
\prod_{i,j}\psi_{\alpha_i\beta_j}(x_i,y_j)\prod_{i}
\phi_{\alpha_i}(x_i)\phi_{\beta_i}(y_i),
\end{equation}
where the pairwise compatibility functions,
$\psi_{\cdot\cdot}(\cdot,\cdot)$, are defined as
\begin{displaymath}
\psi_{\alpha_i\beta_j}(r,s)=\left\{
\begin{array}{cc}
0& r=j \textrm{ and } s\neq i\\
0&r\neq j\textrm{ and } s=i\\
1&\textrm{Otherwise}
\end{array}\right.
\end{displaymath}
the potentials at the nodes, $\phi_{\cdot}(\cdot)$, are defined as
\[
\phi_{\alpha_i}(r)=e^{w_{ir}},~~~\phi_{\beta_j}(r)=e^{w_{rj}},
~~~\forall ~~1\leq ~i,j,r,s~\leq ~n,
\]
and $Z$ is the normalization constant. We note that the pair-wise
potential essentially ensures that the following two constraints are
satisfied for any $(\overline{X}, \overline{Y})$ with positive probability: (a) If
node $\alpha_i$ is matched to node $\beta_j$ (i.e $X_i = j$),
then node $\beta_j$ must be match to node $\alpha_i$ (i.e. $Y_j = i$).
(b) If node $\alpha_i$ is not matched to $\beta_j$ (i.e. $X_i \neq j$),
then node $\beta_j$ must not be matched to node $\alpha_i$ (i.e. $Y_j \neq i$).
These two constraints encode that the support of the above defined
probability distribution is on matchings only.
\begin{claim}
\label{claim:one} For the GM as defined above, the joint density
$p\left(\overline{X}
=(x_1,\dots,x_n),\overline{Y} = (y_1,\dots,y_n)\right)$ is nonzero
if and only if
$\pi_\alpha(\overline{X})=\{(\alpha_1,\beta_{x_1}),(\alpha_2,
\beta_{x_2}),\ldots,(\alpha_n,\beta_{x_n})\}$ and
$\pi_\beta(\overline{Y}) = \{(\alpha_{y_1},\beta_1),(\alpha_{y_2},
\beta_2),\ldots,(\alpha_{y_n},\beta_n)\}$ are both matchings and
$\pi_\alpha(\overline{X})=\pi_\beta(\overline{Y})$. Further, when
nonzero, they are equal to $\frac{1}{Z}e^{2\sum_iw_{ix_i}}$.
\end{claim}
When, $p(\overline{X}, \overline{Y}) > 0$, then the product of
$\phi_{\cdot}(\cdot)$'s essentially make the probability monotone
function of the summation of edge weights as part of the corresponding
matching. Formally, we state the following claim.
\begin{claim} \label{claim:two}
Let $(\overline{X}^*,\overline{Y}^*)$ be such that
$$ (\overline{X}^*,\overline{Y}^*) = \arg\max \{ p\left(\overline{X},
\overline{Y}\right) \}.$$ Then, the corresponding
$\pi_\alpha(\overline{X}^*) = \pi_\beta (\overline{Y}^*)$ is the MWM
in $K_{n,n}$.
\end{claim}

Claim \ref{claim:two} implies that finding the MWM is equivalent to
finding the maximum a posteriori (MAP) assignment on the GM defined
above. Thus, the standard max-product algorithm can be used as an
iterative strategy for finding the MWM. In fact we show that this
strategy yields the correct answer. Before proceeding further, we
provide an example of the above defined GM for the ease of
readability.
\begin{example}
Consider a complete bipartite graph with $n = 2$. The random variables
$X_i, i=1,2$ corresponds to the index of $\beta$ node to which
$\alpha_i$ is connected under the GM. Similarly, the random variable
$Y_i, i=1,2$ correspond to the index of $\alpha$ node to which
$\beta_i$ is connected.  For example, $X_1 = 1$ means
that $\alpha_1$ is connected to $\beta_1$. The pair-wise potential function
$\psi_{\cdot \cdot}$ encodes matching constraints. For example,
$(X_1, X_2; Y_1, Y_2) = (1 2; 1 2)$ corresponds to the matching where $\alpha_1$
is connected to $\beta_1$ and $\alpha_2$ is connected to $\beta_2$. This
is encoded (and allowed) by $\psi_{\cdot \cdot}$: in this example,
$\psi_{\alpha_1 \beta_2}(X_1, Y_2) = \psi_{\alpha_1 \beta_2}(1, 2) = 1$, etc.
On the other hand, $(X_1, X_2; Y_1, Y_2) = (1 2; 2 1)$ is not a matching
as $\alpha_1$ connects to $\beta_1$ while $\beta_1$ connects to $\alpha_2$.
This is imposed by the following: $\psi_{\alpha_1 \beta_1}(X_1, Y_1) = \psi_{\alpha_1 \beta_1}(1, 2) = 0$.
We suggest the reader to go through this example in further detail by him/herself
to get familiar with the above defined GM.
\end{example}

\vspace{.05in}
\subsection{MAX-PRODUCT ALGORITHM FOR $K_{n,n}$}
\vspace{.05in}

Now, we describe the max-product
algorithm (and the equivalent min-sum algorithm) for the GM defined
above.
We need some definitions and notations before we can
describe the max-product algorithm. Consider the following.
\begin{definition} Let $D\in \mathbb{R}^{n\times n}$ and $X,Y,Z\in
\mathbb{R}^{n\times 1}$. Then the operations $\ast,\odot$ are defined
as follows:
\begin{equation} \label{l:op1}
D\ast X = Z\Longleftrightarrow z_i = \max_{j}{d_{ij}x_j}, ~\forall i,
\end{equation}
\begin{equation}\label{l:op2}
X\odot Y = Z \Longleftrightarrow z_i = x_iy_i, ~\forall i.
\end{equation}
For $X_1, \ldots,X_m\in \mathbb{R}^{n\times 1}$,
\begin{equation}\label{l:op3}
\bigodot_{i=1}^m X_i = X_1\odot X_2 \odot \ldots \odot X_m.
\end{equation}
\end{definition}

Define the compatibility matrix
$\Psi_{\alpha_i\beta_j}\in\mathbb{R}^{n\times n}$ such that its
$(r,s)$ entry is $\psi_{\alpha_i\beta_j}(r,s)$, for $1\leq i,j \leq
n$. Also, let $\Phi_{\alpha_i}, \Phi_{\beta_j}\in
\mathbb{R}^{n\times 1}$ be the following:
\[
\Phi_{\alpha_i}=[\phi_{\alpha_i}(1),\ldots,\phi_{\alpha_i}(n)]^t,
\quad \Phi_{\beta_j}=[\phi_{\beta_j}(1),\ldots,\phi_{\beta_j}(n)]^t
\]
where $A^t$ denotes transpose of a matrix $A$.

\vspace{.05in} \noindent{\bf Max-Product Algorithm.} \vspace{.05in}
\hrule \vspace{.1in}
\begin{itemize}
\item[(1)] Let $M_{\alpha_i\rightarrow \beta_j}^k = [m_{\alpha_i\rightarrow\beta_j}^k(1),
m_{\alpha_i\rightarrow\beta_j}^k(2),\ldots,m_{\alpha_i\rightarrow\beta_j}^k(n)]^t\in
\mathbb{R}^{n\times 1}$ denote the messages passed from $\alpha_i$
to $\beta_j$ in the iteration $k \geq 0$, for $1\leq i,j\leq n$.
Similarly, $M_{\beta_j \to \alpha_i}^k$ is the message vector passed
from $\beta_j$ to $\alpha_i$ in the iteration $k$.

\item[(2)] Initially $k = 0$ and set the messages as follows.
Let
$$M_{\alpha_i\rightarrow \beta_j}^0 = [m_{\alpha_i\rightarrow
\beta_j}^0(1) \dots m_{\alpha_i\rightarrow \beta_j}^0(n)]^t, ~~\mbox{and}~~
M_{\beta_j\rightarrow \alpha_i}^0 = [m_{\beta_j\rightarrow
\alpha_i}^0(1) \dots m_{\beta_j\rightarrow \alpha_i}^0(n)]^t, $$
where
\begin{equation} m_{\alpha_i\rightarrow\beta_j}^0(r)  =
 \left\{ \begin{array}{cc} e^{w_{ij}}  & \mbox{~if~} r=i \\
                               1           & \mbox{~otherwise~}
\end{array} \right. \nonumber
\end{equation}
\begin{equation}\label{l:initial}
m_{\beta_i\rightarrow\alpha_j}^0(r)=\left\{
\begin{array}{cc} e^{w_{ji}} & \mbox{~if~} r=i \\
                          1  & \mbox{~otherwise~}
\end{array}\right.
\end{equation}

\item[(3)] For $k \geq 1$, messages in iteration $k$ are obtained from
messages of iteration $k-1$ recursively as follows:
\begin{eqnarray}
M_{\alpha_i\rightarrow\beta_j}^{k} & = &
  \Psi_{\alpha_i\beta_j}^t\ast\Big((\bigodot_{l\neq j}M_{\beta_l\rightarrow
  \alpha_i}^{k-1})\odot\Phi_{\alpha_i}\Big) \nonumber \\
M_{\beta_i\rightarrow\alpha_j}^{k}& = &
\Psi_{\alpha_j\beta_i}\ast\Big((\bigodot_{l\neq
j}M_{\alpha_l\rightarrow\beta_i}^{k-1})\odot\Phi_{\beta_i}\Big)
\label{l:recurse}
\end{eqnarray}

\item[(4)] Define the beliefs ($n\times 1$ vectors) at nodes
$\alpha_i$ and $\beta_j$, $1\leq i,j \leq n$, in iteration $k$ as
follows.
\begin{eqnarray}
b_{\alpha_i}^k&=&\left(\bigodot_{l}M_{\beta_l\rightarrow\alpha_i}^k\right)
\odot\Phi_{\alpha_i} \nonumber \\
b_{\beta_j}^k&=&\left(\bigodot_{l}M_{\alpha_l\rightarrow\beta_j}^k\right)
\odot\Phi_{\beta_j} \label{l:belief}
\end{eqnarray}

\item[(5)] The estimated\footnote{Note that, as defined, $\pi^k$ need
not be a matching. Theorem \ref{thm:one} shows that for
large enough $k$, $\pi^k$ is a matching and corresponds to the MWM.}
MWM at the end of iteration $k$ is $\pi^k$, where
$\pi^k(i) = \arg\max_{1\leq j\leq n} \{ b_{\alpha_i}^k(j) \},$ for
$1\leq i\leq n$.

\item[(6)] Repeat (3)-(5) till $\pi^k$ converges.

\end{itemize}

\vspace{.1in}
\hrule
\vspace{.1in}

\noindent{\bf Note 2.} For computational stability, it is often
recommended that messages be normalized at every iteration.
However, such normalization does not change the output of
the algorithm. Since we are only interested in theoretically analyzing
the algorithm, we will ignore the normalization step. Also, the
messages are usually all initialized to one. Although the result doesn't
depend on the initial values, setting them as defined
above makes the analysis and formulas nicer at the end.

\subsection{MIN-SUM ALGORITHM FOR $K_{n,n}$}\label{s:minsum}
\vspace{.05in}

The max-product and min-sum algorithms can be seen to be equivalent by
observing that the logarithm function is monotone and hence $\max_{i} \log
(\alpha_i) = \log (\max_{i} \alpha_i)$. In order to describe the
min-sum algorithm, we need to redefine $\Phi_{\alpha_i},
\Phi_{\beta_j}$, $1\leq i,j \leq n$, as follows:
\[
\Phi_{\alpha_i}=[w_{i1},\ldots,w_{in}]^t,\quad\Phi_{\beta_j} =
[w_{1j},\ldots,w_{nj}]^t.
\]

Now, the min-sum algorithm is exactly the same as max-product with the
equations (\ref{l:initial}), (\ref{l:recurse}) and (\ref{l:belief})
replaced by:
\begin{itemize}
\item[(a)] Replace (\ref{l:initial}) by the following.
\begin{equation} m_{\alpha_i\rightarrow\beta_j}^0(r)  = \left\{ \begin{array}{cc} {w_{ij}}  & \mbox{~if~} r=i \\
                               0           & \mbox{~otherwise~}
\end{array}\right. \nonumber
\end{equation}
\begin{equation}\label{l:initial1}
m_{\beta_i\rightarrow\alpha_j}^0(r)   = \left\{ \begin{array}{cc} {w_{ji}}  & \mbox{~if~} r=i \\
                                    0          & \mbox{~otherwise~}
\end{array} \right.
\end{equation}

\item[(b)] Replace (\ref{l:recurse}) by the following.
\begin{eqnarray}
M_{\alpha_i\rightarrow\beta_j}^{k}&=&\Psi_{\alpha_i\beta_j}^t\ast\Big((\sum_{l\neq
j}M_{\beta_l\rightarrow\alpha_i}^{k-1})+\Phi_{\alpha_i}\Big) \nonumber \\
M_{\beta_i\rightarrow\alpha_j}^{k}&=&\Psi_{\alpha_j\beta_i}\ast\Big((\sum_{l\neq
j}M_{\alpha_l\rightarrow\beta_i}^{k-1})+\Phi_{\beta_i}\Big)
\label{l:recurse1}
\end{eqnarray}

\item[(c)] Replace (\ref{l:belief}) by the following.
\begin{eqnarray}
b_{\alpha_i}^k & = & (\sum_{l}M_{\beta_l\rightarrow\alpha_i}^k)
+\Phi_{\alpha_i} \nonumber \\
b_{\beta_j}^k&=& (\sum_{l}M_{\alpha_l\rightarrow\beta_j}^k)
+\Phi_{\beta_j} \label{l:belief1}
\end{eqnarray}

\end{itemize}

\vspace{.05in}

\noindent {\bf Note 3.} The min-sum algorithm involves only
summations and subtractions compared to max-product which involves
multiplications and divisions. Computationally, this makes the min-sum
algorithm more efficient and hence very attractive.

\vspace{.05in}
\section{MAIN RESULT}
\label{s:two}
\vspace{.05in}

Now we state and prove Theorem \ref{thm:one}, which is the main
contribution of this paper.
Before proceeding further, we need the following definitions.
\begin{definition} Let $\epsilon$ be the difference between the
weights of the MWM and the second maximum weight matching; i.e.
\[
\epsilon = W_{\pi^*}-\max_{\pi\neq\pi^*}(W_{\pi}).
\]
Due to the uniqueness of the MWM, $\epsilon > 0$.
Also, define $w^* = \max_{i,j}(|w_{ij}|)$.
\end{definition}

\begin{theorem}\label{thm:one} For any weighted complete
bipartite graph $K_{n,n}$ with unique maximum weight matching, the
max-product or min-sum algorithm when applied to the corresponding
GM as defined above, converges to the correct MAP assignment or the
MWM within $\lceil\frac{2nw^*}{\epsilon}\rceil$ iterations.
\end{theorem}

\vspace{.05in}
\subsection{PROOF OF THEOREM \ref{thm:one}}
\vspace{.05in}

\begin{figure}[htpb]
\begin{center}
\epsfig{figure=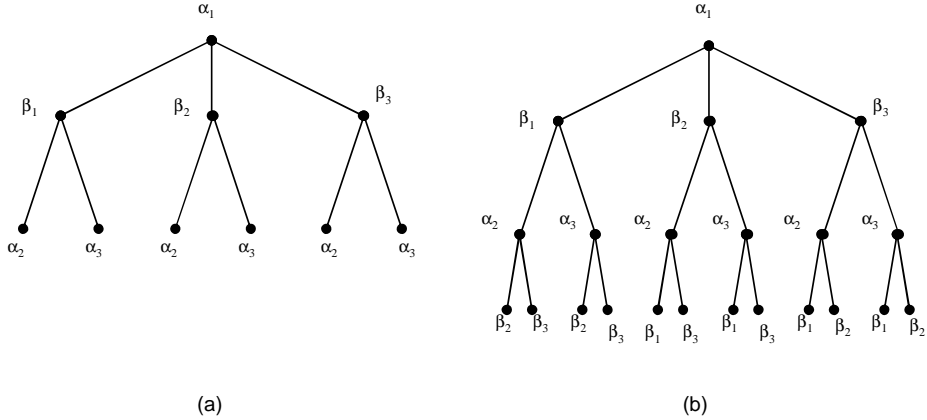,width=0.7\linewidth }
\caption{When $n=3$ (a) is $T_{\alpha_i}^{1}$  and (b) is $T_{\alpha_i}^{2}$.  }
\label{fig:one}
\end{center}
\end{figure}

We first present some useful notation and definitions. Consider
$\alpha_i$, $1\leq i\leq n$. Let $T_{\alpha_i}^k$ be the level-$k$
unrolled tree corresponding to $\alpha_i$, defined as follows:
$T_{\alpha_i}^{k}$ is a weighted regular rooted tree of height $k+1$
with every non-leaf having degree $n$.  All nodes have labels from
the set $\{\alpha_1,\ldots,\alpha_n, \beta_1,\ldots,\beta_n\}$
according to the following recursive rule:~(a) root has label
$\alpha_i$;~(b) the $n$ children of the root $\alpha_i$ have labels
$\beta_1,\ldots,\beta_n$; and (c) the children of each non-leaf node
whose parent has label $\alpha_r$ (or $\beta_r$) have labels
$\beta_1,\ldots,\beta_{r-1},\beta_{r+1},\ldots,\beta_n$ (or $\alpha_1,\ldots,\alpha_{r-1},\alpha_{r+1},\ldots,\alpha_n$). The edge
between nodes labeled $\alpha_i, \beta_j$ in the tree is assigned
weight $w_{ij}$ for $1\leq i,j\leq n$. Examples of such a tree for
$n=3$ are shown in the Figure~\ref{fig:one}.

\vspace{.05in}

\noindent{\bf Note 4.} $T_{\alpha_i}^{k}$ is often called the
level-$k$ {\em computation tree} at node $\alpha_i$ corresponding
to the GM under consideration. The computation tree in general is
constructed by replicating the pairwise compatibility functions
$\psi_{\alpha_i\beta_j}(r,s)$ and potentials
$\phi_{\alpha_i}(r),\phi_{\beta_j}(s)$, while preserving the local
connectivity of the original graph. They are constructed
so that the messages received by node $\alpha_i$ after $k$
iterations in the actual graph are equivalent to those that would
be received by the root $\alpha_i$ in the computation tree, if the
messages are passed up along the tree from the leaves to the root.

A collection $\Lambda$ of edges in computation tree is called a \emph{T-matching} if it no two
edges of $\Lambda$ are adjacent in the tree ($\Lambda$ is a matching in the computation tree) and each
non-leaf nodes are endpoint of exactly one edge from $\Lambda$.  Let $t_{\alpha_i}^k(r)$ be the
weight of maximum weight T-matching in
$T_{\alpha_i}^{k}$ which uses the edge $(\alpha_i,\beta_r)$ at the
root.

Now, we state two important lemmas that will lead to the proof of
Theorem \ref{thm:one}. The first lemma presents an important
characterization of the min-sum algorithm while the second lemma
relates the maximum weight T-matching of the computation tree
and the MWM in $K_{n,n}$.
\begin{lemma}\label{lem:one}
At the end of the $k^{th}$ iteration of the min-sum algorithm, the
belief at node $\alpha_i$ of $K_{n,n}$ is precisely
$b_{\alpha_i}^k=[2t_{\alpha_i}^k(1) \ldots 2t_{\alpha_i}^k(n)]^t$.
\end{lemma}
\begin{lemma}\label{lem:two}
If $\pi^*$ is the MWM of graph $K_{n,n}$ then for
$k>\frac{2nw^*}{\epsilon}$,
$$\pi^*(i) = \arg\max_{r} \{t_{\alpha_i}^k(r)\}.$$
That is, for $k$ large enough, the maximum weight T-matching in
$T_{\alpha_i}^{k}$ chooses the edge $(\alpha_i,\beta_{\pi^*(i)})$ at
the root.
\end{lemma}
\begin{proof}[Theorem \ref{thm:one}]
Consider the min-sum algorithm. Let $b_{\alpha_i}^k =
[b_{\alpha_i}^k(1),\ldots,b_{\alpha_i}^k(n)]^t$. Recall that $\pi^k
= (\pi^k(i))$ where $\pi^k(i) = \arg\max_r \{b_{\alpha_i}^k(r)\}$.
Then, by Lemmas \ref{lem:one} and \ref{lem:two}, for
$k>\frac{2nw^*}{\epsilon}$, $\pi^k = \pi^*$.
\end{proof}
Next, we present the proofs of Lemmas \ref{lem:one} and
\ref{lem:two} in that order.

\begin{proof}[Lemma \ref{lem:one}]
It is known \cite{Weiss1997} that under the min-sum (or max-product)
algorithm, the vector $b_{\alpha_i}^k$ corresponds to the correct
max-marginals for the root $\alpha_i$ of the MAP assignment on the GM
corresponding to $T_{\alpha_i}^k$. The pairwise compatibility
functions force the MAP assignment on this tree to be a T-matching.
Now, each edge has two endpoints and hence its weight is counted
twice in the weight of T-matching.

Next consider the $j^{th}$ entry of $b_{\alpha_i}^k$,
$b_{\alpha_i}^k(j)$. By definition, it corresponds to the MAP
assignment with the value of $\alpha_i$ at the root being $j$. That is,
$(\alpha_i,\beta_j)$ edge is chosen in the tree at the root. From
the above discussion, $b_{\alpha_i}^k(j)$ must be equal to
$2t_{\alpha_i}^k(j)$.
\end{proof}

\begin{proof}[Lemma \ref{lem:two}] Assume the contrary that for some $k>\frac{2nw^*}{\epsilon}$,
\begin{equation}
\pi^*(i) \neq \arg\max_{r}t_{\alpha_i}^k(r) \stackrel{\triangle}{=}
\hat{i}, ~~\mbox{for some $i$}. \label{l:x1}
\end{equation}
Then, let $\hat{i}= \pi^*(i_1)$ for $i_1\neq i$. Let $\Lambda$ be the
T-matching on $T_{\alpha_i}^k$ whose weight is $t_{\alpha_i}^k(\hat{i})$.
We will modify $\Lambda$ and find $\Lambda'$ whose weight is more
than $\Lambda$ and which connects $(\alpha_i,\beta_{\pi^*(i)})$ at the
root instead of $(\alpha_i,\beta_{\pi^*(i_1)})$, thus contradicting
with (\ref{l:x1}).

First note that the set of all edges of $T_{\alpha_i}^k$ whose projection
in $K_{n,n}$ belong to $\pi^*$ is a T-matching which we denote by $\Pi^*$.
Now consider paths $P_{\ell}, ~\ell \geq 0$ in $T_{\alpha_i}^k$, that contain edges from
$\Pi^*$ and
$\Lambda$ alternatively defined as
follows. Let $\alpha_0 = \mbox{root}~\alpha_i$, $i_0 = i$ and $P_1
= (\alpha_0)$ be a single vertex path. Let $P_2 = (\beta_{\pi^*(i_0)},
\alpha_0, \beta_{\pi^*(i_1)})$, where $i_1$ is such that $\alpha_0
= \alpha_i$ is connected to $\beta_{\pi^*(i_1)}$ under $\Lambda$.
For $r \geq 1$, define $P_{2r+1}$ and $P_{2r+2}$ recursively as follows:
\[
P_{2r+1} = (\alpha_{i_{-r}}, P_{2r}, \alpha_{i_r}),
\]
\[
P_{2r+2} = (\beta_{\pi^*(i_{-r})}, P_{2r+1}, \beta_{\pi^*(i_{r+1})})
\]
where $\alpha_{i_{-r}}$ is the node at level $2r$ to which the
endpoint node $\beta_{\pi^*(i_{-r+1})}$ of path $P_{2r}$ is
connected to under $\Lambda$, and $i_{r+1}$ is such that
$\alpha_{i_r}$ at level $2r$ (part of $P_{2r+1}$) is connected to
$\beta_{\pi^*(i_{r+1})}$ under $\Lambda$. Note that, by definition,
such paths $P_{\ell}$ for $\ell \leq k$ exist since the tree
$T_{\alpha_i}^k$ has $k+1$ levels and can support a path of length
at most $2k$ as defined above.

\begin{example} The Figure \ref{fig:two}(d) provides an example
of such a path. The corresponding bipartite graph has $n=3$ with its MWM shown in figure \ref{fig:two}(a). The
Figure \ref{fig:two}(d) shows $T^5_{\alpha_1}$, the computation tree for node
$\alpha_1$, till level $k = 5$. A path, $P_5$, is
highlighted with thick edges alternatively {\em complete and bold} (edges from $\Lambda$) and
{\em dashed} (edges from $\Pi^*$). In the figure, $P_1 = (\alpha_1)$; $P_2 = (\beta_1,\alpha_1,\beta_2)$;
$P_3 = (\alpha_2,\beta_1,\alpha_1,\beta_2,\alpha_2) = (\alpha_3, P_2,\alpha_2)$ and
so on. Finally,
$$P_5 = (\alpha_1,\beta_2,\alpha_2,\beta_1,\alpha_1,\beta_2,\alpha_2,\beta_3,\alpha_3) = C_1 \cup Q,$$
where $C_1 = (\alpha_1,\beta_1,\alpha_2,\beta_2,\alpha_1)$ is a cycle of length $4$ (see
Figure \ref{fig:two}(c)) and $Q =(\alpha_1,\beta_2,\alpha_2,\beta_3,\alpha_3)$ is a path of
length $4$ (see Figure \ref{fig:two}(b)).
\end{example}
\begin{figure}[htbp]
\begin{center}
\begin{tabular}{c}
\end{tabular}
\psfig{figure=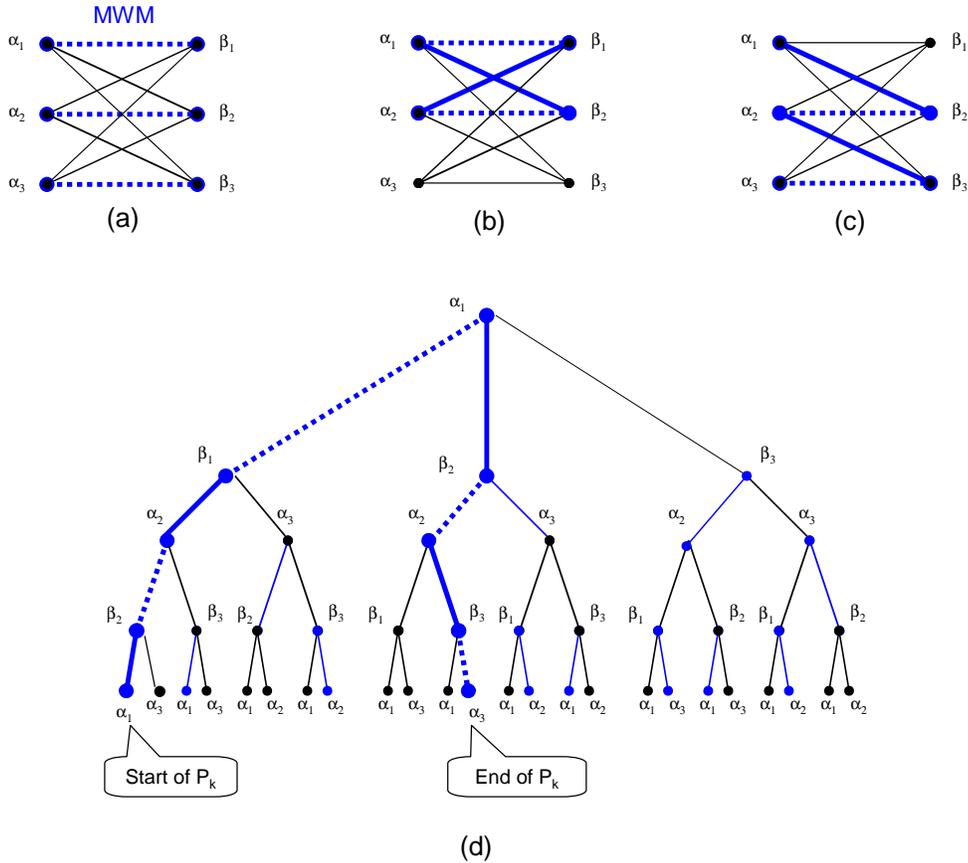,width=.7\linewidth }
\caption{Consider a graph with MWM shown in (a). Projection of the
path $P_k$ for $k=5$ as shown in (d) is decomposed to (b): path Q of
length 4 and (c): cycle $C_1$ of length 4. The {\em dashed} edges
belong to $\Pi^*$ while {\em bold} edges belong to $\Lambda$.}
\label{fig:two}
\end{center}
\end{figure}

Now consider the path $P_k$ of length $2k$. Its edges are alternately
from $\Lambda$ and $\pi^*$. Let us refer to the edges of $\Lambda$ as
the $\Lambda$-edges of $P_k$. Replacing the $\Lambda$-edges of $P_k$ with
their complement in $P_k$ produces a new matching $\Lambda'$ in
$T_{\alpha_i}^k$; this follows from the way the paths are constructed.
\begin{lemma}\label{lem:three}
The weight of T-matching $\Lambda'$ is strictly higher than that of
$\Lambda$ on tree $T_{\alpha_i}^k$.
\end{lemma}
This completes the proof of Lemma \ref{lem:two} since Lemma \ref{lem:three}
shows that $\Lambda$ is not the maximum weight T-matching on $T_{\alpha_i}^k$,
leading to a contradiction.
\end{proof}
Now, we provide the proof of Lemma \ref{lem:three}.

\begin{proof}[Lemma \ref{lem:three}] It suffices to show that the total
weight of the $\Lambda$-edges is less than the total weight of their
complement in $P_k$. Consider the projection $P'_k$ of $P_k$ in the
graph $K_{n,n}$. $P'_k$ can be decomposed into a union of a set of
simple cycles $\{C_1, C_2,\ldots,C_m\}$ and at most one even length
path $Q$ of length at most $2n$. Since each simple cycle has at most
$2n$ vertices and the length of $P_k$ is $2k$,
\begin{eqnarray}
m\geq \frac{2k}{2n} = \frac{k}{n}. \label{l:x1a}
\end{eqnarray}

Consider one of these simple cycles, say $C_s$. Construct the
matching $\pi'$ in $K_{n,n}$ as follows: (i) For $\alpha_l \in C_s$,
select edges incident on $\alpha_l$ that belong to $\Lambda$. Such
edges exist by the property of the path $P_k$ that contains $C_s$.
(ii) For $\alpha_l \notin C_s$, connect it according to $\pi^*$,
that is, add the edge $(\alpha_l, \beta_{\pi^*(l)})$.

Now $\pi' \neq \pi^*$ by construction. Since the MWM is unique,
the definition of $\epsilon$ gives us
\[
W_{\pi'} \leq W_{\pi^*} - \epsilon.
\]
But, $W_{\pi^*}-W_{\pi'}$ is exactly equal to the total weight of
the $\Pi^*$-edges of $C_s$ minus the total weight of the
$\Lambda$-edges of $C_s$. Thus,
\begin{eqnarray}
\textrm{\small weight of $\Lambda$-edges of $C_s$} -
\textrm{\small weight of $\Pi^*$-edges of $C_s$} & =&  -(W_{\pi^*}-W_{\pi'}) ~~\leq~~ -\epsilon.
\label{l:x2}
\end{eqnarray}
Since the path $Q$ is of even length, either the first edge or
the last edge is an $\Lambda$-edge. Without loss of generality,
assume it is the last edge. Then, let
\[
Q =(\beta_{\pi^*(i_{j_1})},\alpha_{i_{j_1}},\beta_{\pi^*(i_{j_2})},\ldots,
\beta_{\pi^*(i_{j_l})},\alpha_{i_{j_l}},\beta_{\pi^*(i_{j_{l+1}})}).
\]
Now consider the cycle
\[
C = (\beta_{\pi^*(i_{j_1})},\alpha_{i_{j_1}},\beta_{\pi^*(i_{j_2})},\ldots,
\beta_{\pi^*(i_{j_l})},\alpha_{i_{j_l}},\beta_{\pi^*(i_{j_1})}).
\]
Alternate edges of $C$ are from the maximum weight matching $\pi^*$.
Hence, using the same argument as above, we obtain
\begin{eqnarray}
\textrm{\small weight of $\Lambda$-edges of $Q$}-\textrm{\small
weight of $\Pi^*$-edges of $Q$} &=&\sum_{1\leq r\leq l}w_{i_{j_r}\pi^*(i_{j_{r+1}})}-\sum_{1\leq
r\leq
l}w_{i_{j_r}\pi^*(i_{j_r})}\nonumber\\
&\leq& -\epsilon + |w_{i_{j_l}\pi^*(i_{j_1})}| +
|w_{i_{j_l}\pi^*(i_{j_{l+1}})}|\nonumber\\
&\leq& -\epsilon + 2w^*. \label{l:x3}
\end{eqnarray}
From (\ref{l:x1a})-(\ref{l:x3}), we obtain that for T-matchings
$\Lambda'$ and $\Lambda$ in $T_{\alpha_i}^k$:
\begin{eqnarray}
\textrm{\small weight of $\Lambda$}-\textrm{\small weight of $\Lambda'$}
&\leq& -(m+1)(\epsilon) + 2w^*\nonumber\\
&\leq& -\frac{k}{n}\epsilon + 2w^*\nonumber\\
&<& 0.
\end{eqnarray}
This completes the proof of Lemma \ref{lem:three}.
\end{proof}

\section{Complexity}\label{s:comp}

In this section, we will analyze the complexity of the min-sum
algorithm described in Section \ref{s:minsum}.  Theorem
\ref{thm:one} suggests that the number of iterations required to
find MWM is $O\left(\frac{nw^*}{\epsilon}\right)$. Now, in each
iteration of Min-Sum algorithm each node sends a vector of
size $n$ (i.e. $n$ numbers)  to each of the $n$ nodes
in the other partition.  Thus, total  number of messages exchanged
in each iteration are $O(n^2)$ with each message of length $n$.
Now, each node performs $O(n)$ basic computational operations
(comparison, addition) to compute each element in a message vector
of size $n$. That is, each node performs $O(n^2)$ computational
operations to compute a message vector in each iteration.  Since
each node sends $n$ message vectors, the total cost is $O(n^3)$
per node or $O(n^4)$ per iteration for all nodes. Thus, total cost
for $O(n w^*/\epsilon)$ iterations is $O(n^5 w^*/\epsilon)$.

Thus, for fixed $w^*$ and $\epsilon$, the running time of algorithm scales as
$O(n^5)$. The known algorithms such as Edmond-Karp's algorithm
\cite{EdmondsKarp1972} or Auction algorithm \cite{Bertsekas} have
complexity of $O(n^3)$.  In what follows, we simplify the Min-Sum
algorithm so that overall running time of the algorithm becomes
$O(n^3)$ for fixed $w^*$ and $\epsilon$. We make a note here that
Edmond-Karp's algorithm is strongly polynomial (i.e. does not depend
on $w^*$ and $\epsilon$) while Auction algorithm's complexity
is $O(n^3 w^*/\epsilon)$.

\subsection{SIMPLIFIED MIN-SUM ALGORITHM FOR $K_{n,n}$}\label{s:minsum1}
\vspace{.05in}

We first present the algorithm and show that it is exactly the same as Min-Sum algorithm.
Later, we analyze the complexity of the algorithm.

\vspace{.05in}
\noindent{\bf Simplified Min-Sum Algorithm.}
\vspace{.05in}
\hrule
\vspace{.1in}
\begin{itemize}
\item[(1)] Unlike Min-Sum algorithm, now each $\alpha_i$ sends
a number to $\beta_j$ and vice-versa. Let the
message from $\alpha_i$ to $\beta_j$ in iteration $k$ be denoted as
$$\hat{m}^k_{\alpha_i\rightarrow \beta_j}$$
Similarly, the
messages from $\beta_j$ to $\alpha_i$ in
iteration $k$ be denoted as
$$\hat{m}^k_{\beta_j \rightarrow\alpha_i}$$

\item[(2)] Initially $k = 0$ and set the messages as follows.
$$ \hat{m}_{\alpha_i\rightarrow\beta_j}^0  = w_{ij}$$
Similarly,
$$ \hat{m}^0_{\beta_j \rightarrow\alpha_i} = w_{ij}$$


\item[(3)] For $k \geq 1$, messages in iteration $k$ are obtained from
messages of iteration $k-1$ recursively as follows:
\begin{eqnarray}
\hat{m}_{\alpha_i\rightarrow\beta_j}^{k} & = & w_{ij} - \max_{\ell \neq j} \hat{m}_{\beta_\ell\rightarrow \alpha_i}^{k-1}, \nonumber \\
\hat{m}_{\beta_j\rightarrow\alpha_i}^{k} & = & w_{ij} - \max_{\ell
\neq i} \hat{m}_{\alpha_\ell\rightarrow \beta_j}^{k-1}
\label{l:recursesim}
\end{eqnarray}

\item[(4)] The estimated MWM at the end of iteration $k$ is $\pi^k$, where
$\pi^k(i) = \arg\max_{1\leq j\leq n}
\{ \hat{m}_{\beta_j\rightarrow\alpha_i}^k \},$ for $1\leq i\leq n$.

\item[(5)] Repeat (3)-(4) till $\pi^k$ converges.

\end{itemize}

\vspace{.1in}
\hrule
\vspace{.1in}

Now, we state and prove the claim that relates the above modified
algorithm to the original Min-Sum algorithm.
\begin{lemma}\label{lem:four}
In Min-Sum algorithm adding an equal amount to all coordinates of
any message vector $M_{\alpha_i\rightarrow\beta_j}^k$ (similarly
$M_{\beta_j\rightarrow\alpha_i}^k$) at anytime does not change the
resulting estimated matching $\pi^m$ for all $k,m$.
\end{lemma}
\begin{proof}
If a number is added to all coordinates of
$M_{\alpha_i\rightarrow\beta_j}^k$ it is not hard to see from
equation (\ref{l:recurse1}) and structure of $\psi_{\alpha_i
\beta_j}(\cdot,\cdot)$ that other message and belief vectors will
change only up to an additive constant to their coordinates. Hence
these changes do not affect $\pi^m(i) = \arg\max_{1\leq j\leq n} \{
b_{\alpha_i}^m(j) \},$ for $1\leq i\leq n$.
\end{proof}
\begin{lemma}\label{lem:bij}
The algorithms Min-Sum and Simplified Min-Sum produce identical
estimated matchings $\pi^m$ at the end of every iteration $m$.
\end{lemma}
\begin{proof}
Consider the Min-Sum algorithm. In particular, consider a message
vector $M_{\alpha_i\rightarrow \beta_j}^k$ in iteration $k$.
First, we claim that all for any given $k \geq 0$,
$m_{\alpha_i\rightarrow \beta_j}^k(r), r \neq i$ are the same.
That is, for $r_1 \neq r_2$ and $r_1, r_2 \neq i$,
$$m_{\alpha_i\rightarrow \beta_j}^k(r_1) = m_{\alpha_i\rightarrow \beta_j}^k(r_2).$$

For $k =0$, this claim holds by definition.  For $k \geq 1$,
consider the definition of $m_{\alpha_i\rightarrow \beta_j}^k(r),
r\neq i$.
\begin{equation}
m_{\alpha_i\rightarrow \beta_j}^k(r) = \max_{1\leq q \leq n}
\psi_{\alpha_i \beta_j}(q,r)
 \left[  w_{iq} + \sum_{\ell \neq j} m_{\beta_\ell \alpha_i}^{k-1}(q)\right]
    = \max_{q \neq j}  \left[  w_{iq} + \sum_{\ell \neq j} m_{\beta_\ell \alpha_i}^{k-1}(q)\right]. \label{l:e1}
\end{equation}
The first equality follows from definition in Min-Sum algorithm
while second equality follows from property of $\psi_{\alpha_i
\beta_j}(\cdot,\cdot)$.  The equation (\ref{l:e1}) is independent of
$r (\neq i)$. This proves the desired claim.

The above stated property of Min-Sum algorithm  immediately implies
that the vector $M_{\alpha_i\rightarrow \beta_j}^k$ has only two
distinct values, one corresponding to $m_{\alpha_i\rightarrow
\beta_j}^k(i)$ and the other corresponding to
$m_{\alpha_i\rightarrow \beta_j}^k(r), r \neq i$. Now subtract
$m_{\alpha_i\rightarrow \beta_j}^k(r), r \neq i$ from all
coordinates of $M_{\alpha_i\rightarrow \beta_j}^k$. Lemma
\ref{lem:four} guarantees the resulting matching $\pi^m$ for all $m$
does not change. Performing the same modification to all message
vectors yields a \emph{Modified Min-Sum} algorithm with the same
outcome as Min-Sum. But each message vector $M_{\alpha_i\rightarrow
\beta_j}^k$ in this Modified Min-Sum has all coordinates equal to
zero except the $i^{th}$ coordinate. Denote these $i^{th}$
coordinates by $\tilde{m}_{\alpha_i\rightarrow \beta_j}^k$. Now
equation (\ref{l:recurse1}) shows these for all $i,j,k$ numbers
$\tilde{m}_{\alpha_i\rightarrow \beta_j}^k$ satisfy the following
recursive equations:
\begin{eqnarray}
\tilde{m}_{\alpha_i\rightarrow\beta_j}^{k} & = & w_{ij} - \max_{\ell \neq j}( \tilde{m}_{\beta_\ell\rightarrow \alpha_i}^{k-1}+w_{i\ell}), \nonumber \\
\tilde{m}_{\beta_j\rightarrow\alpha_i}^{k} & = & w_{ij} - \max_{\ell
\neq i} (\tilde{m}_{\alpha_\ell\rightarrow \beta_j}^{k-1}+w_{\ell
j}) \label{l:recursemod}
\end{eqnarray}
Similarly for new beliefs we have:
\begin{eqnarray}
\tilde{b}_{\alpha_i}^k(r) & = & \tilde{m}_{\beta_r\rightarrow \alpha_i}^k + w_{ir}, \nonumber \\
\tilde{b}_{\beta_j}^k(s) & = & \tilde{m}_{\alpha_s\rightarrow
\beta_j}^k + w_{sj} \label{l:beliefmod}
\end{eqnarray}
Now by adding $w_{ij}$ to each side of (\ref{l:recursemod}) and
dividing them by $2$ it can be seen from (\ref{l:recursesim}) that
numbers $\frac{\tilde{m}_{\alpha_i\rightarrow\beta_j}^{k}+
w_{ij}}{2}$ and $\hat{m}_{\alpha_i\rightarrow\beta_j}^{k}$ satisfy
the same recursive equations. They also satisfy the same initial
conditions. As result for all $i,j,k$ we have
\begin{equation}
\hat{m}_{\alpha_i\rightarrow\beta_j}^{k}=\frac{\tilde{m}_{\alpha_i\rightarrow\beta_j}^{k}+
w_{ij}}{2} =\tilde{b}_{\alpha_i}(j) \label{l:modvssim1}
\end{equation}
and
\begin{equation}
\hat{m}_{\beta_j\rightarrow\alpha_i}^{k}=\frac{\tilde{m}_{\beta_j\rightarrow\alpha_i}^{k}+
w_{ij}}{2} =\tilde{b}_{\beta_j}(i) \label{l:modvssim2}
\end{equation}
This shows that the estimated matching computed at nodes in Modified
Min-Sum and Simplified Min-Sum algorithms are exactly the same at
each iteration which completes the proof of Lemma \ref{lem:bij}.
\end{proof}
\vspace{.05in}

\noindent{\bf Note 5.} The simplified min-sum equations can also be
derived in a direct way by looking interpretation of the messages
$\{\hat{m}_{\alpha_i\to\beta_j}^k\}_{i,j,k}$ in the computation
tree.  More specifically consider the level-$(k+1)$ computation tree
rooted at $\alpha_i$, $T_{\alpha_i}^{k+1}$. Also consider its
subtree, $T_{\alpha_i,\beta_j}^k$ , that is built by adding the edge
$(\alpha_i,\beta_j)$ at the root of $T_{\alpha_i}^{k+1}$ to graph of
all descendants of $\beta_j$. One can show that the message
$\hat{m}_{\beta_j\to\alpha_i}^k$ is equal to the difference between
weight of maximum weight $T$-matching in $T_{\alpha_i,\beta_j}^k$
that uses the edge $(\alpha_i,\beta_j)$ at the root and weight of
the maximum weight $T$-matching in $T_{\alpha_i,\beta_j}^k$ that
does not use that edge. Now a simple induction gives us the update
equations (\ref{l:recursesim}).

\subsection{COMPLEXITY OF SIMPLIFIED MIN-SUM}\label{s:computation}

The Lemma \ref{lem:bij} and Theorem \ref{thm:one} immediately
imply that the Simplified Min-Sum, like Min-Sum, converges after
$O\left(\frac{nw^*}{\epsilon}\right)$ iterations. As described
above, the Simplified Min-Sum algorithm requires total $O(n^2)$
messages per iteration.  Thus, for fixed $w^*$ and $\epsilon$ the
algorithm requires total $O(n^3)$ messages to be exchanged.

Now, we consider the number of computational operations done by each
node in an iteration. From the description of Simplified Min-Sum
algorithm, it may seem that each node will require to do $O(n)$ work
for sending each message and thus $O(n^2)$ work overall at one node.
But, we present a simple method that shows each node can compute
message for all of its $n$ neighbors with $O(n)$ computational
operation (comparison, addition/subtraction). This will result in
$O(n^2)$ overall computation per iteration. Thus, it will take
$O\left(\frac{ n^3w^*}{\epsilon}\right)$ computation in
$O\left(\frac{ nw^*}{\epsilon}\right)$ iterations. This will result
in total complexity of $O\left(\frac{n^3w^*}{\epsilon}\right)$ in
terms of overall messages as well as computation operations.

Here we describe an algorithm to compute messages
$\hat{m}^{k}_{\alpha_1\rightarrow \beta_j}, 1\leq j \leq n$ using
received messages $\hat{m}^{k-1}_{ \beta_j\rightarrow\alpha_1},
1\leq j\leq n$. This is the same algorithm that all $\alpha_i, 1\leq
i\leq n,$ and $\beta_j, 1\leq j \leq n$, need to employ. Now, define
\begin{eqnarray}
i_1 & = & \textrm{argmax}_{1\leq j\leq n}\hat{m}^{k-1}_{
\beta_j\rightarrow\alpha_1}\nonumber\\
i_2 & = & \textrm{argmax}_{1\leq j\leq n,j\neq i_1}\hat{m}^{k-1}_{
\beta_j\rightarrow\alpha_1}\nonumber\\
\Mx_1 & = & \hat{m}^{k-1}_{
\beta_{i_1}\rightarrow\alpha_1}\nonumber\\
\Mx_2 & = & \hat{m}^{k-1}_{ \beta_{i_2}\rightarrow\alpha_1}\nonumber
\label{l:c1}
\end{eqnarray}
Then, from (\ref{l:recursesim}) we obtain
\begin{eqnarray}
\hat{m}^k_{\alpha_1\rightarrow \beta_{i_1}} & = &  w_{1i_1} - \Mx_2, \nonumber \\
\hat{m}^k_{\alpha_1\rightarrow \beta_j} & = &
w_{1j}-\Mx_1\quad\textrm{for $j\neq i_1$}. \label{l:c}
\end{eqnarray}

We see that computing all messages $\hat{m}^k_{\alpha_1\rightarrow
\beta_j}$ takes $O(n)$ operations. From (\ref{l:c}), it takes node
$\alpha_1$ $O(n)$ computations to find $i_1,i_2, \Mx_1, \Mx_2$, then
it takes $O(1)$ computation to compute each of the
$\hat{m}^k_{\alpha_1\rightarrow \beta_j}, 1\leq j\leq n$. That is,
total $O(n)$ operations for computing all messages
$\hat{m}^k_{\alpha_1\rightarrow \beta_j}, 1\leq j\leq n$.

Thus, we have established that each node $\alpha_i, 1\leq i\leq
n,$ and $\beta_j, 1\leq j\leq n,$ need to perform $O(n)$
computation to compute all of its messages in a given iteration.
That is, the total computation cost per iteration is $O(n^2)$. In
summary, Theorem \ref{thm:one}, Lemma \ref{lem:bij} and discussion
of this Section \ref{s:computation} immediately yield the
following result.
\begin{theorem}\label{thm:comp}
The Simplified Min-Sum algorithm finds the Maximum Weight Matching in
$O\left(\frac{nw^*}{\epsilon}\right)$ iterations with total computation cost
of $O\left(\frac{n^3w^*}{\epsilon}\right)$ and $O\left(\frac{n^3w^*}{\epsilon}\right)$
total number of message exchanges.
\end{theorem}


\section{AUCTION AND MIN-SUM}\label{s:auction}

In this section, we will first recall the auction algorithm
\cite{Bertsekas} and then describe its relation to the min-sum
algorithm.

\subsection{AUCTION ALGORITHM FOR MWM}

The Auction algorithm finds the MWM via an ``auction": all
$\alpha_i$ become buyers and  all $\beta_j$ become objects. Let
$p_j$ denote the price of $\beta_j$ and $w_{ij}$ be the value of
object $\beta_j$ for buyer $\alpha_i$. The net benefit of an assignment or
matching $\pi$ is defined as
$$ \sum_{i=1}^n \left(w_{i\pi(i)} - p_{\pi(i)}\right). $$
The goal is to find $\pi^*$ that maximizes this net benefit. It is
clear that for any set of prices $p_1,\ldots,p_n$, the MWM
maximizes the net benefit. The auction algorithm is an iterative
method for finding the optimal prices and an assignment that
maximizes the net benefit (and is therefore the MWM).

\vspace{.05in}
\noindent{\bf Auction Algorithm.}
\vspace{.05in}

\hrule
\vspace{.1in}
\begin{itemize}
\item[$\circ$] Initialize the assignment $S = \emptyset$, the set
of unassigned buyers $I=\{\alpha_1, \ldots,\alpha_n\}$, and prices
$p_j = 0$ for all $j$.

\item[$\circ$] The algorithm runs in two phases, which are
repeated until $S$ is a complete matching.

\item[$\circ$] {\em Phase 1: Bidding.}\\
For all $\alpha_i \in I$,
\begin{itemize}

\item[(1)] Find benefit maximizing $\beta_j$. Let,
\begin{equation}
j_i=\textrm{argmax}_j \{w_{ij}-p_j\}, ~v_i = \max_j\{w_{ij}-p_j\},~ \mbox{and}~ u_i = \max_{j\neq j_i}\{w_{ij}-p_j\}. \label{l:maxben}
\end{equation}

\item[(2)] Compute the ''bid" of buyer $\alpha_i$, denoted by $b_{\alpha_i\to\beta_{j_i}}$ as
follows: given a fixed positive constant $\delta$,
$$b_{\alpha_i\rightarrow\beta_{j_i}}=w_{ij_i}-u_i+\delta.$$
\end{itemize}

\item[$\circ$] {\em Phase 2: Assignment.}\\
For each object $\beta_j$,

\begin{itemize}

\item[(3)] Let $P(j)$ be the set of buyers from which $\beta_j$ received
a bid. If $P(j) \neq \emptyset$, increase $p_j$ to the highest bid,
$$p_j=\max_{\alpha_i\in P(j)}b_{\alpha_i\rightarrow\beta_j}.$$

\item[(4)] Remove the maximum bidder $\alpha_{i_j}$ from $I$ and
add $(\alpha_{i_j}, \beta_j)$ to $S$. If $(\alpha_k, \beta_j) \in
S, ~k\ne i_j$, then put $\alpha_k$ back in $I$.

\end{itemize}
\end{itemize}

\vspace{.1in} \hrule
\vspace{.1in}
\begin{theorem}[\cite{Bertsekas2}] If $0<\delta<\epsilon/n$, then
the assignment $S$ converges to the MWM in $O(n w^*/\epsilon)$
iterations with running time $O(n^3 w^*/\epsilon)$ (where
$\epsilon$ and $w^*$ are as defined earlier).
\end{theorem}

\subsection{CONNECTING MIN-SUM AND AUCTION}

The similarity between equations (\ref{l:c}) and
(\ref{l:maxben}) suggests a connection between
the min-sum and auction algorithms. Next, we describe
modifications to the min-sum and auction algorithms, called {\em
min-sum auction I} and {\em min-sum auction II}, respectively. We
will show that these versions are equivalent and derive some of
their key properties. Here we consider the na\"{i}ve auction algorithm (when
$\delta=0$) and deal with the case $\delta>0$ in the next section.

\vspace{.01in} \noindent{\bf Min-Sum Auction I} \vspace{.05in}

\hrule \vspace{.07in}
\begin{itemize}
\item[(1)] Each $\alpha_i$ sends a number to $\beta_j$ and
vice-versa. Let the messages in iteration $k$ be denoted as
$\tilde{m}^k_{\alpha_i\rightarrow\beta_j},\ \tilde{m}^k_{\beta_j
\rightarrow\alpha_i}\in \mathbb{R}$.

\item[(2)] Initialize $k = 0$ and set
$\tilde{m}_{\beta_j\rightarrow\alpha_i}^0  = 0$.

\item[(3)] For $k \geq 1$, update messages as follows:
\begin{eqnarray}
\tilde{m}_{\alpha_i\rightarrow\beta_j}^{k} & = & w_{ij} - \max_{\ell
\neq j}\{w_{i\ell} - \tilde{m}_{\beta_\ell\rightarrow
\alpha_i}^{k-1}\},
\nonumber\\
\tilde{m}_{\beta_j\rightarrow\alpha_i}^k & = &\max_{\ell=1}^n\
\tilde{m}_{\alpha_\ell\rightarrow \beta_j}^k, \label{l:recursesimauc}
\end{eqnarray}

\item[(4)] The estimated MWM at the end of iteration $k$ is the set of edges
\begin{eqnarray}
\pi^k=\{(\alpha_{i_j},\beta_j)|\ i_j=\arg\max_{1\leq \ell\leq n} \{
\tilde{m}_{\alpha_\ell\rightarrow\beta_j}^k \}\  1\leq j\leq n,
\textrm{ and  } \tilde{m}_{\alpha_{i_j}\rightarrow\beta_j}^k \geq
\tilde{m}_{\beta_j\rightarrow\alpha_i}^{k-1} \}\nonumber
\end{eqnarray}

\item[(5)] Repeat (3)-(4) till $\pi^k$ is a complete matching.

\end{itemize}

\vspace{.02in}\hrule \vspace{.05in}

\vspace{.05in}
\noindent{\bf Min-Sum Auction II.}
\vspace{.1in}

\hrule
\vspace{.1in}
\begin{itemize}
\item[$\circ$] Initialize the assignment $S = \emptyset$ and
prices $p_j = 0$ for all $j$.

\item[$\circ$] The algorithm runs in two phases, which are
repeated until $S$ is a complete matching.

\item[$\circ$] {\em Phase 1: Bidding.}\\
For all $\alpha_i$,

\begin{itemize}

\item[(1)] Find $\beta_j$ that maximizes the benefit. Let,
{\small
\begin{equation}
j_i=\textrm{argmax}_j \{w_{ij}-p_j\}, ~v_i = \max_j\{w_{ij}-p_j\},
~\mbox{and}~u_i = \max_{j\neq j_i}\{w_{ij}-p_j\}. \label{l:maxbena}
\end{equation}}

\item[(2)] Compute the ''bid" of buyer $\alpha_i$, denoted
by $b_{\alpha_i\to\beta_j}$:
$$b_{\alpha_i\rightarrow\beta_{j_i}}=w_{ij_i}-u_i, ~\mbox{and}
~b_{\alpha_i \to \beta_j} = w_{ij} - v_i, ~j \neq j_i.$$
\end{itemize}

\item[$\circ$] {\em Phase 2: Assignment.}\\
For each object $\beta_j$,

\begin{itemize}

\item[(3)] Set price $p_j$ to the highest bid,
$p_j=\max_{\alpha_i}b_{\alpha_i\rightarrow\beta_j}.$

\item[(4)] Reset $S=\emptyset$. Then, for each $j$ add the pair
$(\alpha_{i_j},\beta_j)$ to $S$ if $b_{\alpha_{j_i}\rightarrow
\beta_j} \geq p_j$, where $\alpha_{i_j}$ is a buyer attaining the
maximum in step (3).
\end{itemize}
\end{itemize}

\vspace{.1in} \hrule
\vspace{.1in}

\begin{theorem}\label{thm:msaeq}
The algorithms min-sum auction I and II are equivalent.
\end{theorem}
\begin{proof}
Let $b^k_{\alpha_i \to \beta_j}$ and $p^k_j$ denote the
bids and prices at the end of iteration $k$ in algorithm min-sum auction II.
Now, identify $b^k_{\alpha_i \to \beta_j}$ with $\tilde{m}^k_{\alpha_i\to \beta_j}$ and
$p^k_j$ with $\tilde{m}^k_{\beta_j\to \alpha_i}$. Then it is immediate
that min-sum auction II becomes identical to min-sum auction I. This completes
the proof of Theorem \ref{thm:msaeq}.
\end{proof}

Next we will prove that if the min-sum auction algorithm
terminates (we omit reference to I or II), it finds the correct
maximum weight matching. As we will see, the proof uses standard
arguments (see \cite{Bertsekas} for example).
\begin{theorem}\label{thm:msaterm}
Let $\sigma$ be the termination matching of the min-sum auction I
(or II). Then it is the MWM, i.e. $\sigma = \pi^*$.
\end{theorem}
\begin{proof}
The proof follows by establishing that at termination, the
messages of min-sum auction form the optimal solution for the dual
of the MWM problem and $\sigma$ is the corresponding optimal
solution to the primal, i.e. MWM. To do so, we first state the
dual of the MWM problem
\begin{eqnarray}
 \min & & \sum_{i=1}^n r_i + \sum_{j=1}^n p_j \nonumber \\
 \mbox{subject to} ~& & r_i + p_j ~\geq~ w_{ij}. \label{ed0}
\end{eqnarray}
Let $(r^*, p^*)$ be the optimal solution to the above stated dual
problem and let $\pi^*$ solve the primal MWM problem. Then, the
standard complimentary slackness conditions are:
\begin{eqnarray}
r^*_i + p^*_{\pi^*(i)} & = & w_{i\pi^*(i)}. \label{ecs}
\end{eqnarray}
Thus, $(r^*,p^*, \pi^*)$ are the optimal dual-primal solution for
the MWM problem if and only if (a) $\pi^*$ is a matching, (b)
$(r^*,p^*)$ satisfy (\ref{ed0}), and (c) the triple satisfies
(\ref{ecs}). To complete the proof we will prove the existence of
$r^*, p^*$ such that $(r^*, p^*, \sigma)$ satisfy (a), (b) and
(c).

To this end, first note that $\sigma$ is a matching by the
termination condition of the algorithm; thus, condition (a)is
satisfied. We'll consider the min-sum auction II algorithm for the
purpose of the proof. Suppose the algorithm terminates at some
iteration $k$. Let $p^{k-1}_j$ and $p^k_j$ be the prices of
$\beta_j$ in iterations $k-1$ and $k$ respectively. Since all
$\beta_j$s are matched at the termination, from step (4) of the
min-sum auction II, we obtain
\begin{eqnarray}
p^{k}_j & \geq & p^{k-1}_j, ~~\forall j. \label{ef0}
\end{eqnarray}
At termination (iteration $k$), $\alpha_i$ is matched with
$\beta_{\sigma(i)}$ or $\beta_j$ is matched with
$\alpha_{\sigma^{-1}(j)}$. By the definition of the min-sum
auction II algorithm,
\begin{eqnarray}
p^k_j & = & w_{\sigma^{-1}(j) j} - \max_{\ell \neq j}
\left[  w_{\sigma^{-1}(j) \ell} - p^{k-1}_\ell  \right]. \label{ef1}
\end{eqnarray}
From (\ref{ef0}) and (\ref{ef1}), we obtain that
\begin{eqnarray}
w_{\sigma^{-1}(j) j} - p^k_j & \geq & \max_{\ell \neq j} \left[
w_{\sigma^{-1}(j) \ell} - p^{k}_\ell  \right]. \label{ef2}
\end{eqnarray}
Define, $r^*_i = w_{i \sigma(i)} - p^k_{\sigma(i)}$ and $p^*_j =
p^k_j$. Then, from (\ref{ef2}) $(r^*, p^*)$ satisfy the dual
feasibility, that is (\ref{ed0}). Further, by definition they
satisfy the complimentary slackness condition (\ref{ecs}). Thus,
the triple $(r^*, p^*, \sigma)$ satisfies (a), (b) and (c) as
required. Hence, the algorithm min-sum auction II produces the
MWM, i.e. $\sigma = \pi^*$. §
\end{proof}

The min-sum auction II algorithm looks very similar to the auction
algorithm and inherits some of its properties. However, it also
inherits some properties of the min-sum algorithm. This causes it
to behave differently from the auction algorithm. The proof of
convergence of auction algorithm relies on two properties of the
auctioning mechanism: (a) the prices are always non-decreasing and
(b) the number of matched objects is always non-decreasing. By design, (a)
and (b) can be shown to hold for the auction algorithm. However,
it is not clear if (a) and (b) are true for min-sum auction.
In what follows, we state the result that prices are
eventually non-decreasing in the min-sum auction algorithm; however
it seems difficult to establish a statement similar to (b) for the
min-sum algorithm as of now.
\begin{theorem}\label{thm:priceinc}
If $\pi^*$ is unique then in the min-sum auction II algorithm,
prices eventually increase. That is, $\forall k\in
\mathbb{Z}^+;~\exists~T > k~ s.t.~\forall t\geq
T;~p_j^t>p_j^k,~1\leq j\leq n$
\end{theorem}
\begin{proof}
Proof of Theorem
(\ref{thm:priceinc}) is essentially based on (i)
the equivalence between the min-sum auction algorithms I and II,
and (ii) arguments very similar to the ones used in the proof of
Lemma \ref{lem:two} , where we relate prices with the
computation tree.
\end{proof}

Our simulations suggests that in the absence of the condition
``$\tilde{m}_{\alpha_{i_j}\rightarrow\beta_j}^k \geq \tilde{m}_{
\beta_j\rightarrow\alpha_i}^{k-1}$" from step (4) of min-sum auction
I, the algorithm always terminates and finds the MWM as long as it
is unique. This along with Theorem \ref{thm:priceinc} leads us to
the following conjecture.

\begin{conjecture}\label{conj:main} If $\pi^*$ is unique then the
min-sum auction I terminates in a finite number of iterations if
condition ``$\tilde{m}_{\alpha_{i_j}\rightarrow\beta_j}^k \geq
\tilde{m}_{ \beta_j\rightarrow\alpha_i}^{k-1}$" is removed from step
(4).
\end{conjecture}

\subsection{RELATION TO $\delta$-RELAXATION}

In the previous section, we established a relation between the
min-sum and auction (with $\delta=0$) algorithms. In
\cite{Bertsekas, Bertsekas2} the author extends the auction
algorithm to obtain guaranteed convergence in a finite number of
iterations via a $\delta$-relaxation for some $\delta > 0$. At
termination the $\delta$-relaxed algorithm produces a triple
$(r^*,p^*, \pi^*)$ such that (a1) $\pi^*$ is a matching, (b1)
$(r^*, p^*)$ satisfy (\ref{ed0}) and (c1) the following modified
complimentary slackness conditions are satisfied:
\begin{eqnarray}
r^*_i + p^*_{\pi^*(i)} & \leq & w_{i\pi^*(i)} + \delta. \label{ecs1}
\end{eqnarray}
The conditions (c1) are referred to as $\delta$-CS conditions in
\cite{Bertsekas}. This modification is reflected in the description
of the auction algorithm where we have added $\delta$ to each bid in
step (2). We established the relation between min-sum and auction
for $\delta = 0$ in the previous section. Here we make a note that
for every $\delta > 0$, the similar relation holds. To see this, we
consider min-sum auction I and II where the bid computation is
modified as follows: modify step (3) of min-sum auction I as $
\tilde{m}^k_{\alpha_i \to\beta_j} = w_{ij} - \max_{\ell \neq j}
\{w_{i\ell} - \tilde{m}_{\beta_\ell\rightarrow \alpha_i}^{k-1}\} +
\delta, $ and modify step (2) of min-sum auction II as
$b_{\alpha_i\rightarrow\beta_{j_i}}=w_{ij_i}-u_i+\delta, ~\mbox{and}
~b_{\alpha_i \to \beta_j} = w_{ij} - v_i + \delta, ~j \neq j_i.$
For these modified algorithms, we obtain the following result
using arguments very similar to the ones used in Theorem
\ref{thm:msaterm}.
\begin{theorem}\label{thm:msaterm1}
For $\delta > 0$, let $\sigma$ be the matching obtained from the
modified min-sum auction algorithm I (or II). Then, $w_\sigma \geq
w_{\pi^*} - n \delta$ (i.e. $\sigma$ is within $n\delta$ of the
MWM).
\end{theorem}

%
%
%
%
%
%

\section{DISCUSSION AND CONCLUSION}\label{s:conc}

In this paper, we proved that the max-product algorithm converges
to the desirable fixed point in the context of finding the MWM for
a bipartite graph, even in the presence of loops. This result has
a twofold impact. First, it will possibly open avenues for a
demystification of the max-product algorithm. Second, the same
approach may provably work for other combinatorial optimization
problems and possibly lead to better algorithms.

Using the regularity of the structure of the problem, we managed
to simplify the max-product algorithm. In the simplified algorithm
each node needs to perform $O(n)$ addition-subtraction operations
in each iteration. Since $O(n)$ iterations are required in the
worst case, for finite $w^*$ and $\epsilon$, the algorithm
requires $O(n^3)$ operations at the most. This is comparable with
the best known MWM algorithm. Furthermore, the distributed nature
of the max-product algorithm makes it particularly suitable for
networking applications like switch scheduling where scalability
is a necessary property.

Future work will consist of trying to extend our result to finding
the MWM in a general graph, as our current arguments do not carry
over\footnote{A key fact in the proof of lemma \ref{lem:three} was
the property that bipartite graphs do not have odd cycles.}. Also,
we would like to obtain tighter bounds on the running time of the
algorithm since simulation studies show that the algorithm runs much
faster on average than the worst case bound obtained in this paper.


\section*{Acknowledgment}

While working on this paper 
D. Shah was supported
by NSF grant CNS  - 0546590.




%

\end{document}